\documentstyle[12pt,aps,epsfig,epsf,floats]{revtex}
\tighten
\begin{document}
\title{Nuclear Teleportation }
\author
{$^{\dagger}$\,B.\,F. Kostenko\,\thanks{e-mail: kostenko@cv.jinr.ru}
\and $^{\dagger}$\,V.\,D. Kuznetsov
\and $^{\ddagger}$\,M.\,B. Miller
\and \mbox{$^{\ddagger}$\,A.\,V. Sermyagin}\and \mbox{and $^{\dagger}$\,D.\,V. Kamanin}}
\address{$^{\dagger}$ Joint Institute for Nuclear Research\\%
Dubna Moscow Region 141980, Russia\\
$^{\ddagger}$Institute of Physical and Technology Problems\\ P.O.\,Box 39 Dubna Moscow
Region 141980, Russia\\}
\maketitle
\begin{abstract}
Until recently, only science-fiction authors ventured to use a term teleportation. However,
in the last few years, on the eve of upcoming new millennium, the situation changed very much. The present report
gives a synopsis of main concepts in this area. The readers will be able to make sure that paradoxical
phenomena in the microcosm give a possibility to demonstrate the exchange of properties between microobjects,
removed at a very large distance from each other, when no forces act between them. A new experimental scheme
with hydrogen and helium nuclei is proposed. It is expected that the results of these experiments will be
considered as
teleportation of
nuclear properties of atoms of the simplest chemical elements. A problem of teleportation of
the  more palpable
cargo is left to the physics of the more distant future.
\smallskip
\end{abstract}

\subsection*{Introduction}

It was in the middle of  twenties that an analysis of transportation of soya beans on
the Chinese Eastern Railway was carried out. It appeared that counter transportation
constituted a greater share of the total cargo traffic. Then an original procedure of processing
 the cargoes was invented: in the number of cases it was possible to deliver bean lots
 to recipients from the nearest stations, where at that time there was a sufficient amount
 of beans of a corresponding category, intended, though, to be sent to some other and more
 remote points. Economy of a rolling stock and other advantages for the railway were obvious.
  The history fails to mention how this innovation ended. Probably the complicated events
  on the CER in the beginning of the thirties put an end to the promising experiment.
  Nevertheless,  this was perhaps a first attempt to realize the supertransportation of dry
  substances, or particulate solids.

\bigskip
The process of teleportation (commonly accepted term for supertransportation) according to
usual understanding is reduced to moving through space in such a way that
the object to be transported disappears at one spot of space and reappears exactly at the
same time in some other point. It is well understood that it is not necessary to move
through the space the matter the object is composed of. It is enough to extract
an exact information about inner properties of the object, then transmit this information
to a predetermined  place, and use it afterward to reconstruct the initial
object from a stuff that comes to hand at the point of destination. Thus the teleportation
results in disappearing of the object with its initial properties in the initial place and
 the identical object to reappear in another place. Without disappearing it would not be the
 teleportation, but merely a reproduction, i.\,e. a creation of a new identical specimen, or
  a copy of the object. Let us look how physicists cope with this problem.
\subsection*{Action-at-a-distance (teleporting information?)}
In 1935 Albert Einstein and his colleagues Boris Podolsky and Nathan Rosen (EPR) developed a gedanken
experiment to show as they thought a defect in quantum mechanics (QM)\,\cite{EPR,Bohr}. This
experiment has obtained the
name of the EPR-paradox, and essence of the paradox is as follows.
There are two particles that interacted with each other for some time and have constituted a
single
system. Within the framework of QM that system is described by a certain wave function.
When the
 interaction of the particles is finished and they flew far away from
 each other,
 these two particles are still described by the same wave function as
  before. However, individual states
 of each separate
  particle are completely unknown, moreover, definite individual properties   do not exist in principle
  as quantum mechanics postulates dictate.
   It is only after one of the particles is registered by a particle-detection system that the states
   arise to existence
   for both particles. Furthermore, these states are generated instantly and simulteneously regardless of
   the distance
between the particles at the moment. This scheme is used to be considered sometimes as
teleportation of information possible at a speed higher than that of light.
The real (not only "gedanken") experiments on teleportation of
information, in the sense of EPR-effect, or "a spooky-action-at-a-distance", as A. Einstein
called it,
were carried out only 30-35 years later, in the
seventies-eighties\,\cite{Aspect,Clauser}. Experimenters, however, managed to achieve full
and
definite success only with photons (quanta of visible light),
though, experiments with atoms\,\cite{atoms} and protons (nuclei of hydrogen)
were also performed\,\cite{proton}.
For the case of photons, the experiments were carried out for various distances between
the members of the EPR-pairs in the moment of registration.
The EPR-correlation between the complementary photons was shown to survive
up to as large distances as  more than ten kilometers from one to another
photon\,\cite{10 kilometers}.
In the case of protons, the experiment was carried out only for much
 smaller distances (of about a
few centimeters) and a condition of so-called causal separation, $\Delta x>c\Delta t$,
was not met. Thus, it was not fully convincing, as have been recognized by the
authors of the work themselves.

\subsection*{Teleporting photon-quantum state (or the light
quantum itself?)}

A next step in this way that suggested itself was  not merely
 "action-at-a-distance",
 but the teleportation at least of  a quantum state
from one quantum
 object to another. In spite of the successful experiments with the net
 EPR-effect, it was
 thought until
  recently  that even this kind of teleportation is at best
  a long way in the
  future, if at all.
At first sight it  seems that the Heisenberg uncertainty principle forbids the
first necessary step of the teleportation
procedure: the extraction of complete information about the inner properties of the quantum
object. This is because of the impossibility to obtain simultaneously the exact
values of so-called complementary
variables of a quantum microscopic object (e.\,g., spatial coordinates and momenta).
Nevertheless, in 1993, a group of physicists (C.\,Bennet and his colleagues) managed
to get round this
difficulty\,\cite{Bennett}.
 They showed that full quantum information is not necessary for the process of transferring
 quantum states from one
 object to another which are at an arbitrary large distance from each other. Besides,
 they proposed that
  a
 so-called EPR-channel of communication has to be created on the basis
 of the EPR-pair of two
 quantum object
 (let it be photons B and C, shown in FIG.\,1).
\begin{figure}[h]
\includegraphics[width=\textwidth]{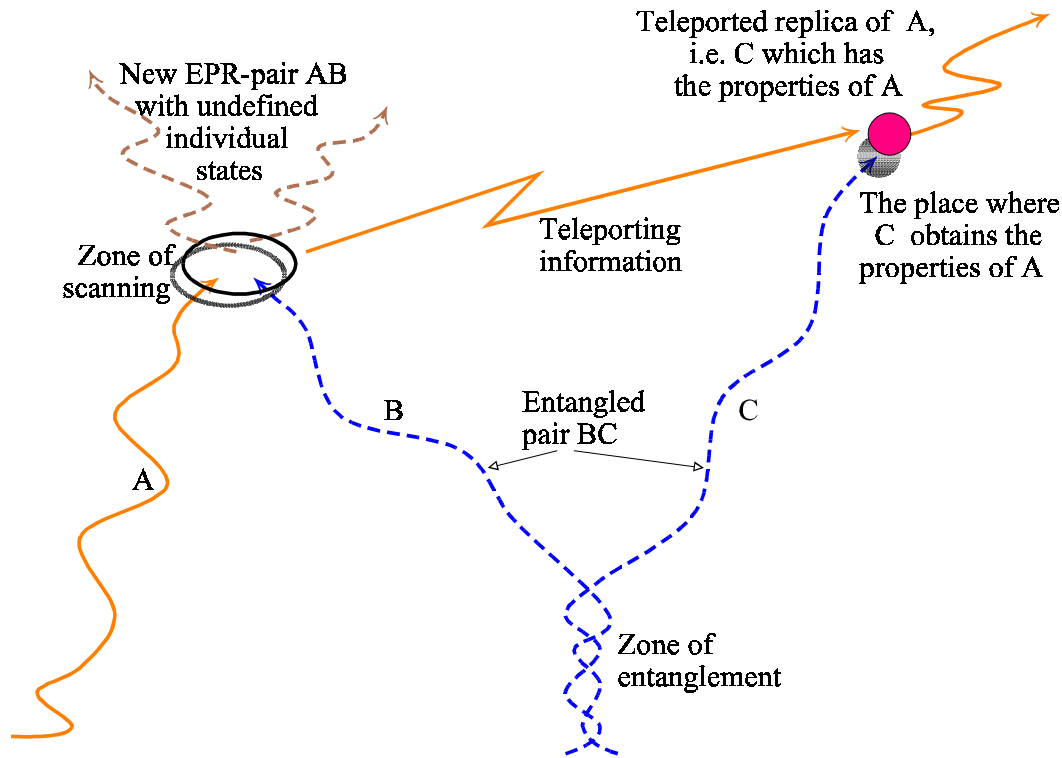}
\caption{Illustration of a general idea of how the teleportation can be realized. Here A is
a light photon we want to pass to a destination place,  B and C, representing an EPR-pair
 of photons, constitute
a so-called quantum transmission channel. As a result,
definite properties of A are destroyed completely at the zone of scanning, and at another place
 we
have photon with  the properties A  had just before it met intermediary object B ("vehicle").
Note
that the vehicle first contacts with (so to say "visits") the C-photon to which the "cargo" has to be
transported, and only later it calls A to take the cargo from it!}
\label{cntrate}
\end{figure}
 After they have interacted in a way to form a
 single system
 decaying afterward,
 the photon B is directed to a "point of departure", where it meets A in a device
 (a registration
 system) arranged in a mode to "catch" only those events, in which B
 will appear in the state, leaving no choice to its "EPR-mate" but
 to take the state A had
 initially -- before the
 interaction with B in the detector at the "point of departure".
 This experimental technique is very fine but well known to
 those skilled in the EPR-art. The conservation laws of general physics are the basis of the
 procedure realizing the system with a given selective sensitivity.
 A result of all these manipulations is that particle C gets
 something from A. It is only the quantum state. Unfortunately, not a soya bean,
 but all the same it is something.
What is important from the point of view of QM, is the
disappearing of A in the place, notified in  FIG.\,1 as a "Zone
of scanning" (ZS). That is, the procedure of interaction of B and A
photons destroyed A photon, in a sense that of two photons outgoing from
the ZS no one has definite properties of A. They constitute a new
EPR-pair of photons, which only as a whole has the definite quantum
state, the individual components of the pair are deprived of these
properties. Thus, the photon A disappears at ZS. Exactly at the
same moment the photon C obtains the properties A had in the
beginning. Once it has happened, in view of the principle of identity
of elementary particles, we can say
that A, disappearing at ZS, reappears at another place, i.\,e., the
teleportation is accomplished.
 This process has several paradoxical features. In spite of the absence of
  contacts between objects (particles, photons) A and C, A manages to pass its properties to C.
  It may be arranged in such a way,
  that
  the distance from A to C is large enough to prevent any exchange of signals between A and C.
  And last,
  but not
  least of interest, in contrast to the transportation of ordinary material cargo,
  when a delivery vehicle
  first visits
  the
  sender to collect the cargo from it, in the case of cargo as subtle as quantum properties,
  it is delivered in a backward fashion. Here the photon
 B plays a role of the delivery vehicle, and we can see
  that
  B first visits (interacts with) the recipient (photon C) and only after that it travels
  to the sender (A)
  for the cargo.

 Finally, to
 reconstruct
initial object completely, it is necessary to fix a time moment when
the interaction of A and B
occurred
(the moment of the arrival of the "vehicle" to the departure "station"
after it visits the
recipient), and
accomplish the required
experimental
data
 processing in due manner. The task of recording the moment of (A-B)-interaction
 and using it in the data analysis together with the information transmitted by a quantum
 EPR-channel
 requires
 one more channel of communication, an ordinary or classical
  transmission line. Receiving information that A and B to form a
  new EPR--pair (using a classical telecommunication line), an observer
  in the point of destination may be sure that the properties of C
  are identical to those of A before the teleportation.

\bigskip
The new idea was immediately recognized as extremely important and a
 few  groups of
experimenters
set
 forth concurrently to implement it. Nevertheless, it took more than four years to overcome
  all technology obstacles
 in the way to realize the project\,\cite{Zeilinger,DBOSCHI}.
 This is because every experiment in this
  field, being a record by itself, is always one  step farther beyond the limits of experimental state of the art
  achieved before.

\subsection*{Start with protons}

An analysis of the problem carried out by authors of the present
experimental project
 which is now in a stage of preparation
takes them to a conclusion
that the experimental setups and instruments developed for usual, though the most
modern,
nuclear-physics studies
(high-current  accelerators
of protons and heavier nuclei, liquid\cite{Liquid} and polarized\cite{Polarized} hydrogen
targets,
 multi-parameter near $2\pi$-geometry -- i.\,e. semi-spherical
 \mbox{aperture --} facility for
particle detection named
"Fobos" at Flerov Laboratory of Nuclear Reaction  of the Joint Institute for
Nuclear Research\cite{FOBOS}),
allow one to design
 a new way to perform the teleportation of the "heavy" matter
 (i.\,e., with
 non-zero mass at rest), with
  prospects to realize the project
in a short time.
Thus, the teleportation of the protons
(nucleus of hydrogen atoms) could be achieved in about a year, and it would
take about two years to prepare the teleportation of more heavy nuclei,
e.g., $^3$He. The concept of
  measurements consists in recording signals entering two independent
but strictly synchronized memory devices with the aim to select
afterwards only those events  that for sure appeared to be
causally separate, for even the most rapid signal (light) could not
connect them.
\begin{figure}[h]
\includegraphics[width=\textwidth]{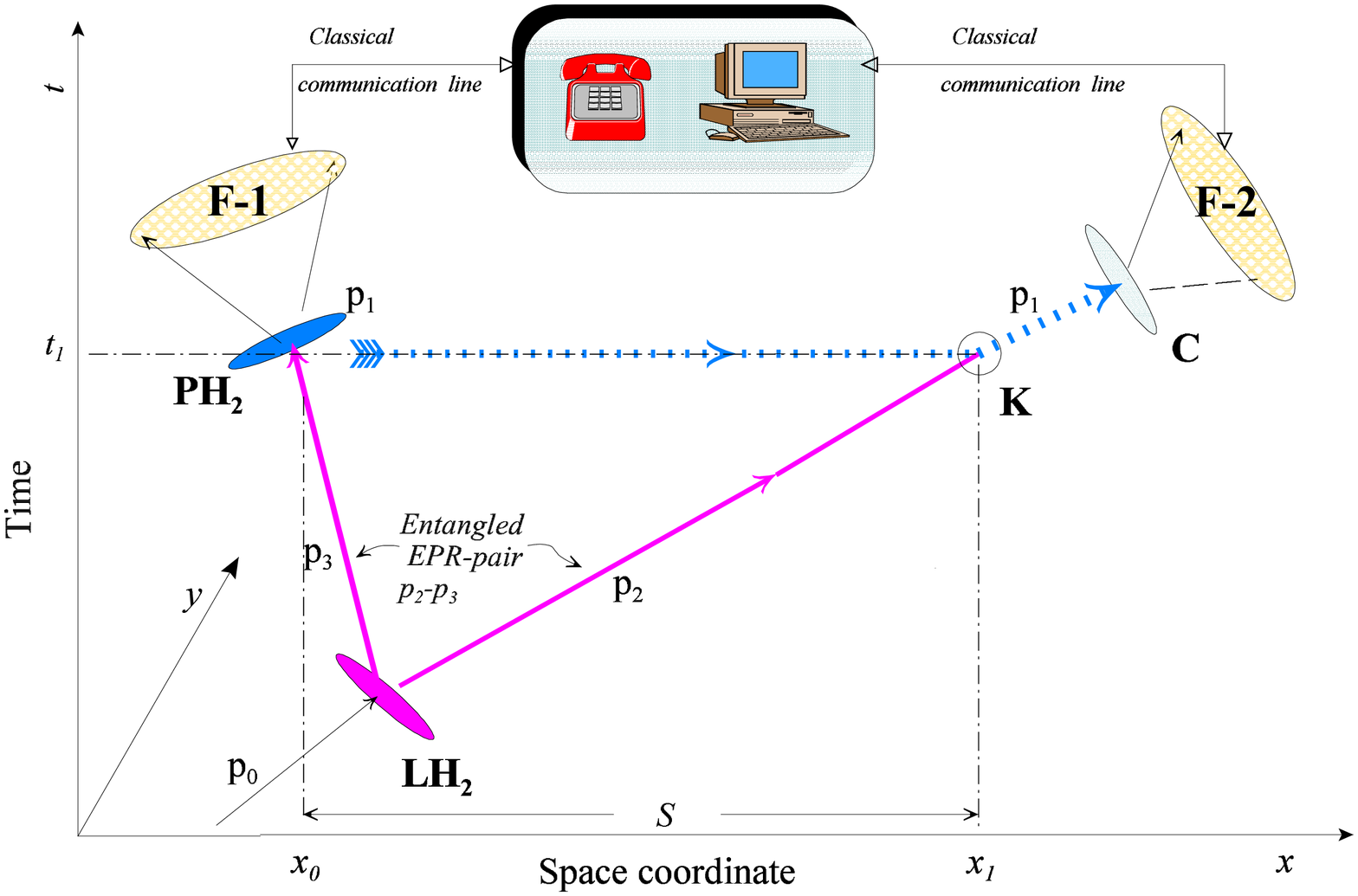}
\caption{Layout of the experiments on proton teleportation. p$_0$ is an
initial proton from
 the accelerator,
LH$_2$ - liquid hydrogen target, p${_2}$p${_3}$ - entangled EPR-pair, PH$_2$ - polarized
 hydrogen target,
C - carbon target
operating as an analyzer of the polarization of protons by a sign of scattering angle (left-right asymmetry),
 F-1 and F-2 - large-aperture position-sensitive particle detectors (so-called Fobos-facilities). Proton
  spin-state is being teleported from the PH$_2$-target placed at x$_0$ to the point x$_1$.
  It can be arranged
  that no
  signal from x$_0$ has enough time to reach point x$_1$ before p$_2$ obtains properties
   of p$_1$ at a
   moment t$_1$. That fact is justified by the detection system F-1/F-2 connected with
   a data-processing center by
   usual communication lines. K is a point, where the spin of p${_2}$ gets a definite
   orientation: just the same,
   that  one of the protons p$_1$ in the PH$_2$-target had
   before the scattering of p${_3}$ from it; the proton p$_1$ loses its definite quantum
   state, as it forms   a new
   EPR-pair together with the scattered proton  p${_3}$.}
\label{prod}
\end{figure}
FIG.\,2  shows the layout of the experiment on teleportation of spin states of
protons from a
polarized target PH$_2$ into the point of destination (target C). A proton beam p$_0$ of
a suitable
energy within the 20-50 MeV range
 bombards the liquid-hydrogen target LH$_2$. According to
 the known experimental data, the scattering in
the target LH$_2$ in the direction of a second target (i.\,e., at the c.m. angle
 $\theta\simeq 90^\circ$)
within a few percent occurs through a
so-called singlet intermediary state, characterized by a zero total spin of the
two-proton system\,\cite{proton}.
Thus, the outgoing  p$_2$ and p$_3$ protons present a two-proton
entangled
system and
are fully analogous to the EPR-correlated photons
used for transmitting
information via the quantum communication channel in the experiments
on the teleportation of
"massless" matter
(light photons), as it was discussed in the preceding section. One of
the scattered protons,
p$_2$, then travels to the point of destination
(target-analyzer C), while the other,
p$_3$, comes to a point where the teleportation is expected to be
started, i.\,e., to the
PH$_2$-target. The latter is used as  a source of particles we are going to teleport.
In this sense, protons within this
target play the same role as photons A in the above section. There
are two features differentiating the case of protons from that of photons.
First, protons p$_1$ are within the motionless target (and, thus, they
are motionless themselves) where their density is greater; besides,
the protons within the PH$_2$-target have quite a
definite quantum state, determined by a direction of polarization.
The last circumstance allows one to perform the experiment under
controllable conditions, i.\,e., this gives the possibility to check the
expected result of the teleportation action.
 In the case
when the scattering in the
polarized target PH$_2$ occurs under the same kinematics conditions
as in the target
LH$_2$ (i.\,e., at the c.m. angle
$\theta\simeq 90^\circ$),
the total spin of the particles p$_1$
and p$_3$  must also be equal to zero after collision.
To detect these events,
a removable circular module F-1 of the facility "Fobos" is supposed to
be used, thus, the
detection
efficiency
is hoped to be much enhanced. According to QM, if all the above conditions
are provided,
the
protons
reaching a point K suddenly receive the same spin projections as the protons in
the
polarized target PH$_2$ have. Therefore, the teleportation of the spin
states from
 the PH$_2$-target
to the recipient p$_2$ really takes place at the point K.
Thus, if the coincidence mode of the detection is provided via any
classical channel, then a
 strong
correlation has to take place between polarization direction in the
target
PH$_2$ and the
direction of the
deflection of p$_2$-protons scattered in the carbon target C. C plays a
 role of the
analyzer of polarization: the protons are deflected to the left or to
the right depending on
sign of their polarization, i.\,e., the orientation of the proton
spin that
can have only two alternatives (along or opposite to a given direction
\cite{analyzer}).
The second module  of "Fobos", designated F-2 in the FIG.\,2, crowns the procedure of
teleportation, as it indicates the proton scattering direction in the
carbon target C, and hence,
 its
polarization.

If we succeeded to make a distance between
the detectors F-1 and F-2 to be sufficiently large, then it would be
possible to meet the important criteria of the space-like interval
(causal independence) between the events of the "departure" of the
quantum state from the
PH$_2$-target and "arrival" of this "cargo" to the recipient
(p$_2$-proton) at the point K. To prevent any exchange of signals
between the points PH$_2$ and K, it is essential to choose appropriate
proportions of some time and space segments, indicated in FIG.2.
Namely, we have to obtain $S>ct_{12}$, where $t_{12} =|t_{F1}-t_{F2}|$.
Here $t_{F1}$ and $t_{F2}$ are moments of registration of signals
from the corresponding detectors F-1 and F-2 (their arrival at the
data collection-processing center). For simplicity, we neglected
a time of flight of the protons from K to
C, and from the PH$_2$- and C-targets to the detectors F-1 and F-2,
respectively.

\subsection*{Conclusion}
Finally, referring to the principle of identity of elementary particles
of the same kind with
the same
quantum characteristics, i.\,e. the protons in our case, we can say that protons from a
polarized target PH$_2$ are
transmitted to the destination point C (through the point K). Thus,
in the nearest future, teleportation of
protons can  come from the domain of dreams and fiction to the
reality in physicists' laboratories.

\smallskip

Remembering that
the above soybeans contains not only protons but as well proteins,
somebody
perhaps feels
disillusioned.
However, we should not be stingy, something should be left for physics
of the third
millennium.

\bigskip

The work was supported in part by  the
Russian Foundation for Basic Research, projects nr. 99-01-01101.

\newpage

\newpage

\end{document}